\documentclass[aps,prb,preprint,superscriptaddress,reprint,preprintnumbers,amssymb,longbibliography]{revtex4-1}

\usepackage{amsmath}
\usepackage{amssymb}
\usepackage{braket}
\usepackage{bbold}
\usepackage[version=3]{mhchem}
\usepackage{graphicx}
\usepackage{color}
\usepackage{units}
\usepackage{pifont}
 \usepackage[caption=false]{subfig}
\usepackage{multirow}
\usepackage[dvipsnames]{xcolor}
\usepackage{hyperref}
\usepackage[final]{changes}

\DeclareMathOperator{\Tr}{Tr}

\AtBeginDocument{\usepackage{booktabs}}

\definechangesauthor[name=Stephan Mohr, color=red]{SM}
\definechangesauthor[name={Stephan Mohr}, color=ForestGreen]{LG}
\definechangesauthor[name={Stephan Mohr}, color=blue]{LR}
\definechangesauthor[name={Stephan Mohr}, color=magenta]{MM}
\definechangesauthor[name={done}, color=black]{done}

\newcommand{\good}{\textcolor{ForestGreen}{\ding{52}}}
\newcommand{\bad}{\textcolor{red}{\ding{56}}}

\newcommand{\tablesize}{\small}

\begin{document}

\author{Stephan Mohr}
\affiliation{Barcelona Supercomputing Center (BSC)}
\email{stephan.mohr@bsc.es}

\author{Michel Masella}
\affiliation{Laboratoire de Biologie Structurale et Radiologie, Service de Bio\'energ\'etique, Biologie Structurale et M\'ecanisme,Institut de Biologie et de Technologie de Saclay, CEA Saclay, F-91191 Gif-sur-Yvette Cedex, France}

\author{Laura E. Ratcliff}
\affiliation{Argonne Leadership Computing Facility, Argonne National Laboratory, Illinois 60439, USA}
\affiliation{Now at Department of Materials, Imperial College London, SW7 2AZ, United Kingdom}

\author{Luigi Genovese}
\affiliation{Univ.\ Grenoble Alpes, INAC-MEM, L\_Sim, F-38000 Grenoble, France}
\affiliation{CEA, INAC-MEM, L\_Sim, F-38000 Grenoble, France}
\email{luigi.genovese@cea.fr}

\title{Complexity Reduction in Large Quantum Systems: Reliable Electrostatic Embedding for Multiscale Approaches via Optimized Minimal Basis Functions}


\date{\today}

\begin{abstract}
Given a partition of a large system into an active quantum mechanical (QM) region and its environment, we present a simple way of embedding the QM region into an effective electrostatic potential representing the environment.
This potential is generated by partitioning the environment into well defined fragments, and assigning each one a set of electrostatic multipoles, which can then be used to build up the electrostatic potential.
We show that, providing the fragments and the projection scheme for the multipoles are chosen properly, this leads to 
an effective electrostatic embedding of the active QM region which is
of equal quality as a full QM calculation.
We coupled our formalism to the DFT code \textsc{BigDFT}, which 
uses a minimal set of localized in-situ optimized basis functions;
this property eases the fragment definition while still describing the electronic structure with great precision.
Thanks to the linear scaling capabilities of \textsc{BigDFT}, we can compare the modeling of the electrostatic embedding with results coming from unbiased full QM calculations of the entire system.
This enables a reliable and controllable setup of an effective coarse-graining approach, coupling together different levels of description, which yields a considerable reduction in the degrees of freedom and thus
paves the way towards efficient QM/QM and QM/MM methods for the treatment of very large systems.
\end{abstract}

 
\maketitle

\section{Introduction}
Density Functional Theory (DFT) calculations based on the Kohn-Sham formalism~\cite{hohenberg-inhomogeneous-1964,kohn-self_consistent-1965} allow the treatment of considerably larger systems than other first principles approaches.
However the intrinsic cubic scaling of this method still limits the size of the systems which can be modeled to typically some hundred atoms.
With the introduction of linear scaling algorithms~\cite{goedecker-linear-1999,bowler-O(N)-2012} this limit can be considerably extended, and calculations up to several thousand atoms can nowadays be done in a routine way~\cite{ratcliff-2016-challenges}.
Nevertheless there are situations where it is desirable --- either due to performance issues or for conceptual considerations --- to apply an effective complexity reduction (ECR), meaning that parts of the system
are treated at a coarser level of theory.
In this context a natural approach is to ``split'' the system into an active quantum mechanical (QM) region, which is treated at a full QM level of theory, and an environment, which is treated at lower computational cost.
From now on, we will denote such a setup as an \emph{embedding} of the QM region into the environment.


Similar in spirit, but still conceptually different are the various ECR methods based on a \emph{fragmentation} of a big QM system.
Here the system is not only split into an active region and an environment, but rather partitioned into several fragments, which are each individually treated on a strict ab-initio level, but which are mutually interacting in a simplified way.
The various ECR methods based on such a fragmentation usually only differ in the way in which they deal with the problem of describing the mutual interactions between the fragments.

One of the most popular methods for such a simplified interaction between the fragments of a large QM system is the Fragment Molecular Orbital (FMO) approach~\cite{kitaura-fragment-1999,fedorov-extending-2007}, which assigns the electrons to some user-defined fragments and then solves the electronic structure problem for each fragment, taking into account the electrostatic potential generated by the other fragments.
Very similar in spirit is the X-Pol method~\cite{gao-toward-1997,gao-a-molecular-1998,wierzchowski-hydrogen-2003,xie-design-2007,xie-the-variational-2008,xie-incorporation-2008,wang-multilevel-2012,gao-variational-2012}, where each one of the fragments is again treated with electronic structure theory, whilst the interactions among them are handled by a combined quantum mechanics/molecular mechanics (QM/MM) approach.
Another method is the Molecular Tailoring Approach (MTA)~\cite{gadre-molecular-1994,ganesh-molecular-2006,gadre-molecular-2006,sahu-molecular-2014}, where a large system is decomposed --- either manually or automatically, depending on geometrical criteria and chemical intuition --- into small overlapping fragments, for which individual QM calculations are performed and eventually combined to get the desired quantities for the entire system.
Finally we also mention subsystem DFT~\cite{jacob-subsystem-2014,krishtal-subsystem-2015},
which is conceptually different 
as it is based on a fragmentation of the electronic charge density and not of the atoms of the system.
Moreover, within this approach an effective embedding potential --- which is exact under certain conditions, i.e.\ it is equivalent to a traditional Kohn-Sham DFT calculation of the entire system --- arises in a natural way.
All of these fragmentation approaches allow quasi-first-principles calculations up to the order of one million atoms~\cite{nishimoto-density-functional-2014}, and therefore offer interesting advantages for the quantum mechanical treatment of large-scale systems.

Nevertheless, these \emph{fragmentation} methods are conceived to simplify the \emph{full} ab-initio calculation of a big QM system, i.e.\ they aim at treating the entire system at the same level of theory.
This is in contrast to \emph{embedding} approaches, which use various levels of theory within a single calculation, thus paving the way towards coarse grain models which can be used within multiscale QM/MM simulations,
able to tackle systems that are orders of magnitude larger than those accessible with traditional quantum approaches~\cite{ratcliff-2016-challenges}.
Among others, we quote here the QM/MM methods detailed in Refs.~\citenum{bakowies-hybrid-1996, ESPF_2002, lin-QM/MM-2007,senn-QM/MM-2009}.

Obviously the introduction of an ECR raises the question of the pertinence of this approximation, and it would be desirable to dispose of an effective measurement of how the given ECR affects the ab-initio character of the calculation.
In the context of ECRs based on embedding, 
the accuracy and reliability 
depends on two factors:
Firstly on how the splitting into the QM region and environment is done, and secondly on the way in which the environment and the interactions with the QM region are treated.


In this paper we will focus on the quality of the embedding potential,
proposing a simple framework for an effective electrostatic embedding of the active QM region into an environment.
Since it is an embedding and not a fragmentation method, we clearly distinguish between the active QM region and the embedding (environmental) region, 
and thus use different levels of theory for the regions.
This clear distinction allows us to focus on the essential parts of the system and thus to use the available computational resources in an efficient way.

Our embedding potential is based on a multipole expansion of the electrostatic potential of the environment.
To derive these multipoles,
we come back to the idea of partitioning the entire systems into fragments.
To each of those well defined fragments we assign a set of partial and static electrostatic charges, like those playing a key role in standard MM force fields (see among others Ref.~\citenum{mackerell-empirical-2004}),
which can also be used to quantify charge transfer effects among important moieties of a molecular system and even to get inputs for particular kinds of quantum mechanical methods, like constrained DFT calculations~\cite{dederichs-ground-1984,wu-direct-2005,kaduk-constrained-2012,ratcliff-fragment-2015}.
These electrostatic charges --- possibly enhanced by higher multipoles --- of each subsystem can then be employed to define an electrostatic environment where the QM active region can be embedded.

Obviously the quality of the electrostatic embedding is directly related to the quality of the fragment multipoles.
Deriving a general and reliable way of partitioning an arbitrary system into a set of fragments is a very complex task.
However, in a previous publication we derived a simple way of determining in a quantitative way whether a chosen fragmentation is reasonable, and whether the associated multipoles can be considered as meaningful ``pseudo-observables'' with an interpretable physico-chemical meaning~\cite{mohr-fragments1-2017}.
In this paper we now demonstrate that fragments which fulfill this condition lead to an accurate embedding potential that reaches the same quality as a full QM treatment of the entire system.

The determination of the correct fragmentation and calculation of the associated multipoles is based on a full QM calculation of the entire system and thus does not directly provide any speedup compared to a straightforward full QM approach.
However, our method allows us to reliably address the problem of \emph{defining} an electrostatic embedding which is consistent with the unbiased full QM calculation.
In other terms, we can use the full QM calculations to validate \emph{a posteriori} whether an embedding based on environment fragments is meaningful or not.
Consequently, the existence of a linear scaling approach which is capable of treating large systems is crucial for this validation procedure.
Fortunately, the \textsc{BigDFT} code~\cite{genovese-daubechies-2008} into which we implemented our embedding procedure exhibits such a linear scaling mode~\cite{mohr-daubechies-2014,mohr-accurate-2015}, thus enabling us to perform calculations involving thousands of atoms in relatively little time.
Once this validation is done, we can then use the embedding framework to perform computationally more demanding simulations, in this way accessing new quantities which would otherwise be out of reach.

The outline of this paper is as follows.
In Sec.~\ref{sec: Subsystem representation} we show how we can create from a set of fragment multipoles an electrostatic potential into which a QM system can be embedded.
In Sec.~\ref{sec: applications} we then show various examples to validate and apply this embedding approach, which can then be used for forthcoming applications.
In Sec.~\ref{sec: pentacene} we demonstrate that the behavior of a large bulk system can be exactly reproduced, and show how the embedding allows one to use expensive hybrid functionals which would be out of reach for a straightforward full QM calculation;
and in Sec.~\ref{sec: DNA} we then discuss in detail the correct setup for a solvation embedding of a large complex system, demonstrating the necessity of including a small shell of explicit solvent.

\section{Subsystem representation via the electrostatic potential}
\label{sec: Subsystem representation}

\subsection{Fragmentation identification and multipole assignment}
\label{sec: Fragmentation identification and multipole assignment}
The goal of this work is to show how a set of (pseudo-)observable fragment multipoles can be used for an electrostatic embedding of a QM region and thus lead to an ECR.
We presented in Ref.~\citenum{mohr-fragments1-2017} a detailed discussion of how the fragments can be identified and the set of associated multipoles calculated; we provide here for consistency a brief summary of the most important results.

A complete description of a QM system is given by the density matrix operator $\hat{F}$, as the measurement of any observable $\hat{O}$ can be written as $\Tr(\hat{F}\hat{O})$.
For KS-DFT, where the many-body wave function $\ket{\Psi}$ giving rise to $\hat{F}$ is given by a single Slater determinant, the density matrix is idempotent, i.e.\ $\hat{F}^2=\hat{F}$.
For a QM system which is separable into reasonably ``independent'' fragments, we want to define, in an analogous way, a fragment density matrix $\hat{F}^{\mathfrak{F}}$, such that a fragment observable can then be evaluated as $\Tr(\hat{F}^{\mathfrak{F}}\hat{O})$.
In such a case, it is reasonable to assume that the fragment density matrix can be written as
\begin{equation}
 \hat{F}^\mathfrak{F} = \hat{F}\hat{W}^\mathfrak{F} \;,
\end{equation}
where $\hat{W}^\mathfrak{F}$ is a projection operator onto the fragment.

To proceed further, we assume that the density matrix  and the fragment projector can both  be written in terms of a set of localized basis functions $\ket{\phi_\alpha}$:
\begin{align}
 \hat{F} &= \sum_{\alpha,\beta} \ket{\phi_\alpha} K_{\alpha\beta} \bra{\phi_\beta} \;,\\
 \hat{W}^\mathfrak{F} &= \sum_{\mu,\nu} \ket{\phi_\mu} R_{\mu\nu}^\mathfrak{F} \bra{\phi_\nu} \;, 
\end{align}
where the $\ket{\phi_\alpha}$ are from now on called \textit{support functions}, the matrix $\mathbf{K}$ the \textit{kernel}, and the matrix $\mathbf{R}$ defines the character of the projection.
In this way the above mentioned idempotency translates in a purity condition for the kernel, i.e.\ $\mathbf{K}\mathbf{S}\mathbf{K} = \mathbf{K}$, with $S_{\alpha\beta} = \braket{\phi_\alpha|\phi_\beta}$.

If the fragment were indeed an independent subsystem, the associated density matrix should also be idempotent, i.e.\ $(\hat{F}^\mathfrak{F})^2 = \hat{F}^\mathfrak{F}$.
For improper fragment definitions this condition will however be violated.
In order to measure the ``fragment property'' of a subsystem, we thus consider the so-called \textit{purity indicator}, defined by
\begin{equation}
 \begin{aligned}
  \Pi &= \frac{1}{q} \Tr\left( \left(\hat{F}^\mathfrak{F} \right)^2 - \hat{F}^\mathfrak{F} \right) \\
      &= \frac{1}{q} \Tr\left( \left(\mathbf{ K S}^\mathfrak{F} \right)^2 - \mathbf{ K S}^\mathfrak{F}\right) \;,
 \end{aligned}
 \label{eq:normalized_purity_indicator}
\end{equation}
where $q$ is the total number of electrons of the isolated fragment in gas phase and $\mathbf S^\mathfrak{F}\equiv \mathbf S \mathbf R^\mathfrak{F} \mathbf S$.


In what follows, we are interested in characterizing the fragments via their electrostatic multipole moments.
Given a charge density $\rho(\mathbf r)$, its multipole moments are given by the expression
\begin{multline} \label{basicqlm}
 Q^R_{\ell m} \equiv  \sqrt{\frac{4\pi}{2\ell + 1}} \int \mathcal{S}_{\ell m}(\mathbf r - \mathbf{r}_R) \rho(\mathbf r) \, \mathrm d \mathbf r \\ = 
 \sqrt{\frac{4\pi}{2\ell + 1}} \Tr \left( \hat F \hat{\mathcal{S}}^R_{\ell m} \right) =
 \Tr\left(\mathbf K \mathbf{P}^R_{\ell m} \right)
 \;,
\end{multline}
where we have defined the multipole matrices $\mathbf{P}^R_{\ell m}$ as
\begin{equation}
P^R_{\ell m;\alpha\beta} = \sqrt{\frac{4\pi}{2\ell+1}}\braket{\phi_\alpha|\hat {\mathcal{S}}^R_{\ell m}|\phi_\beta} \;,
 \label{eq:basic_equation_atomic_multipole_matrices}
\end{equation}
and $\hat {\mathcal{S}}^R_{\ell m}(\mathbf r) \equiv \mathcal{S}_{\ell m}(\mathbf r - \mathbf{r}_R)$ are the solid harmonic operators centered on the reference position $\mathbf r_R$.
As is explained in more detail in Ref.~\citenum{mohr-fragments1-2017}, the electrostatic multipoles of a fragment can be written in a similar way as
\begin{equation} \label{fragmult}
 Q^\mathfrak{F}_{\ell m} \equiv 
 \Tr(\mathbf{K}\mathbf{S}\mathbf{R}^\mathfrak{F}\mathbf{P}^\mathfrak{F}_{\ell m}) \;.
\end{equation}

A popular choice for the fragments are the individual atoms, leading to atomic multipoles $Q^A_{\ell m}$.
In this case the above definition can be seen as a generalization of the well-known Mulliken and L\"owdin charge population analyses, 
and we can associate specific projector matrices to these traditional approaches, as is demonstrated in Ref.~\citenum{mohr-fragments1-2017}.

If we are interested in the multipole moments for a fragment which is composed of various sub-fragments, 
the projector $\hat W^\mathfrak{F}$ onto the large fragment can simply be defined as the sum over the projectors onto the small fragments.
Typically these sub-fragments are the individual atoms, and we thus get $\hat W^\mathfrak{F}=\sum_{A \in \mathfrak{F}} \hat W^A$.
We may then combine the atomic multipoles to obtain the corresponding quantities for the 
fragments, as is shown in more detail in Ref.~\citenum{mohr-fragments1-2017}.
For the important case of the monopole and dipole of a fragment the result is very simple and given by
\begin{subequations}
\begin{align}
  Q^{\mathfrak{F}}_{00} &= \sum_A Q_{00}^A \;, \label{molmonopole} \\ 
  Q^{\mathfrak{F}}_{1 m} &= \sum_A  \sqrt{\frac{3}{4\pi}}\mathcal{S}_{1m}(\mathbf{r}_\mathfrak{F}-\mathbf{r}_A) Q_{00}^A + Q_{1m}^A \; . \label{moldipole}
 \end{align}
\end{subequations}

\subsection{Charge analysis schemes}
\label{sec: Charge analysis schemes}
Suppose that a fragment $\mathfrak F$ has been identified in a QM calculation.
If we are not interested in the QM information of this fragment (which is the typical case for solvent or environment molecules),
we might \emph{reduce} the complexity of the  calculation by expressing only the electrostatic potential generated by this fragment,
thereby lowering the number of atoms within the QM region and consequently also the computational cost.
To do so we choose to represent the electrostatic potential via a set of multipoles.
Before describing this approach in more detail, we want to give a quick overview over popular schemes for the determination of the set of multipoles.

Loosely speaking the various (atomic)
population analyses can be divided into three main classes: i) approaches based on a set of localized atom-centered orbitals,  ii) grid-based methods working directly with the electronic charge density, and iii) methods which determine the atomic charges by fitting them to the electrostatic potential. 

With respect to the first class, the most renowned examples are the Mulliken charge population analysis~\cite{mulliken-electronic-1955}, the L\"owdin population analysis~\cite{loewdin-on-1950,loewdin-on-1970}, and approaches like the natural population analysis (NPA)~\cite{reed-natural-1985}.
The convenience of all the methods belonging to this first class is that they work in the subspace of the atomic orbitals where the density matrix of the system is represented.

Methods belonging to the second class work directly with the charge density represented on a numerical grid and try to partition it into disjunctive regions, which are then assigned to the individual atoms. One of the most popular is Bader's atoms-in-molecule approach~\cite{bader-quantum-1981,bader-1991-a-quantum}, which defines the boundary between two atoms as the 2D surface through which the charge density has ``zero flux'', meaning that it exhibits a local minimum there. We also quote here the Voronoi deformation density approach~\cite{fonseca-voronoi-2004} that assigns the charge of each grid point to the closest atom, taking into account only geometrical information and neglecting the nature of the atoms. This is in contrast to the Hirshfeld approach~\cite{hirshfeld-bonded-atoms-1977}, which partitions the charge density on each grid point to the surrounding atoms according to a weight function which is based on the atomic charge densities, in this way taking into account the nature of the atoms.
The shortcoming of these approaches is that they can be very sensitive with respect to the resolution of the grid, and moreover they require a large amount of data, making them potentially very expensive. An even more important drawback is that they are based on purely geometrical criteria to partition the system. Such a  partitioning scheme may be doubtful from a QM perspective as only the electrostatic charge distribution is taken into account.

The methods belonging to the third class differ mainly in the way the fitting procedure~\cite{sigfridsson-comparison-1998} is performed to reproduce the electrostatic properties of a molecular system on the nodes of a surrounding grid.
Popular approaches are the CHELP method~\cite{chirlian-atomic-1987}, CHELPG~\cite{breneman-determining-1990} and the 
Merz-Singh-Kollman scheme~\cite{singh-an-approach-1984,besler-atomic-1990}. We also mention here
the approach derived by Bayly et al.~\cite{bayly-a_well-behaved-1993}, where the charges are restrained using a penalty function  to avoid unreasonably large charges resulting from the fit, as well as the ESPF method\cite{ESPF_2002}. 
In these methods, the original QM information is encoded in a local function (the electrostatic potential), which is identified beforehand. By construction the QM partitioning is thus imposed by the procedure, thereby leaving the appropriateness of the choice of the fragments to the chemical intuition of the user.
As a side note, we mention that there exist also approaches which assign charges to locations other than atomic centers, e.g.\ chemical bonds~\cite{aleman-multicentric-1994}, to get a more accurate description.

In our approach we use a method based on the density matrix of the full system, as briefly summarized in Sec.~\ref{sec: Fragmentation identification and multipole assignment}.
As is shown in Ref.~\citenum{mohr-fragments1-2017}, a minimal set of in-situ optimized basis functions allows the use of conceptually simple methods such as Mulliken or L\"owdin while still obtaining reliable and unbiased results.
In addition our multipoles are evaluated \emph{a posteriori}, based on a full QM calculation of the entire system, in this way eliminating any bias potentially caused by a inappropriate choice of the fragments.
Moreover the purity indicator of Eq.~\eqref{eq:normalized_purity_indicator} allows one to determine whether the multipoles can be interpreted as physically meaningful ``pseudo-observables''.

\begin{table*}[t]\tablesize
 \centering
\begin{tabular}{l l rr r rr r rr r rr}
 \toprule
  && \multicolumn{2}{c}{\ce{H2O}} && \multicolumn{2}{c}{\ce{HF}} && \multicolumn{2}{c}{\ce{H2} -- atomic} && \multicolumn{2}{c}{\ce{H2} -- bond}\\
  \cmidrule{3-4} \cmidrule{6-7} \cmidrule{9-10} \cmidrule{12-13}
    \multicolumn{1}{c}{threshold} && \multicolumn{1}{c}{volume} & \multicolumn{1}{c}{error} && \multicolumn{1}{c}{volume} & \multicolumn{1}{c}{error} && \multicolumn{1}{c}{volume} & \multicolumn{1}{c}{error} && \multicolumn{1}{c}{volume} & \multicolumn{1}{c}{error} \\
    $10^{-4}$  && 97.0\% & 3.4\% && 97.0\% & 2.4\% && 97.6\% & 95.3\% && 97.6\% & 0.9\% \\
    $10^{-6}$  && 94.3\% & 3.0\% && 93.7\% & 2.1\% && 95.1\% & 95.4\% && 95.1\% & 0.5\% \\
    $10^{-8}$  && 93.3\% & 2.9\% && 92.5\% & 2.1\% && 94.3\% & 95.4\% && 94.3\% & 0.5\% \\
    $10^{-10}$ && 91.9\% & 2.8\% && 90.9\% & 2.0\% && 93.0\% & 95.4\% && 93.0\% & 0.5\% \\
    $10^{-12}$ && 90.5\% & 2.7\% && 89.2\% & 2.0\% && 91.7\% & 95.4\% && 91.7\% & 0.5\% \\
  \bottomrule
\end{tabular}
 \caption{
 Relative difference between the exact electrostatic potential and its multipole approximation, according to Eq.~\eqref{eq:relative_error_potential}. 
 Since the long range behavior shall be analyzed, this quantity is only calculated for grid points where the charge density is below a given threshold. Additionally we indicate the relative amount of the simulation volume which is covered by each threshold. For \ce{H2} we report two results, namely firstly the standard setup where the atomic multipoles are placed on the atoms, 
 and secondly a modified setup where the multipoles are placed on the covalent bond.
}
 \label{tab:molecule_potential_difference}
\end{table*}

\subsection{Representation of the electrostatic potential}
\label{sec: Representation of the electrostatic potential}
Given the definition of the fragment $\mathfrak{F}$ and the associated set of multipoles, we now detail the representation of the electrostatic potential.
We choose to express the fragment's charge density 
by a sum of localized functions based on atom-centered Gaussians:
\begin{equation}
 \begin{gathered}
  \rho^\mathfrak F(\mathbf{r}) =\sum_{A \in \mathfrak F} \rho^A(\mathbf r)\;, \\
  \rho^A(\mathbf r) \simeq  e^{-\frac{|\mathbf r -\mathbf r_A|^2}{2\sigma^2}} \sum_{\ell,m} a_\ell Q_{\ell m}^A  S_{\ell m}^A(\mathbf r) \;,
 \end{gathered}
\end{equation}
where the $Q_{\ell m}^A$ are the atomic multipoles of the atoms forming the fragment $\mathfrak{F}$, and the coefficients $a_\ell$ are defined in Appendix~\ref{sec:Approximating the electrostatic density from a set of multipole}.
This expression guarantees the preservation of the original values of the fragment multipoles $Q_{\ell m}^\mathfrak{F}$. 
The spread $\sigma$ of each Gaussian is an arbitrary quantity that is associated with the ``atomic density'' within the fragment.
In order to represent the long-range electrostatic properties of the fragment, it is enough to set this value close to the characteristic extension of the core electron density, in our case specified by the employed pseudopotentials~\cite{hartwigsen-relativistic-1998,willand-norm-conserving-2013}.
The potential $V(\mathbf{r})^A$ can then be expressed by solving Poisson's equation,
\begin{equation}
 \nabla^2 V^A(\mathbf{r}) = -4 \pi \rho^A (\mathbf{r}) \; ;
 \label{eq:potential_from_Poisson_equation}
\end{equation}
using our Poisson solver based on interpolating scaling functions~\cite{genovese-efficient-2006,genovese-efficient-2007,cerioni-efficient-2012,fisicaro-a-generalized-2016} this calculation is possible for any kind of boundary conditions, making this approach very flexible.
When the Gaussian functions can be considered as a point-charge distribution 
(for example for fragments located far from the simulation domain),
the potential can be directly calculated using the analytic formula for the multipole expansion of an electrostatic potential,
 which reads, up to quadrupoles,
\begin{equation}
 V^A(\mathbf{r}) = - \frac{Q_{00}^A}{|\mathbf{r}-\mathbf{R}^A|} - \frac{\mathbf{r}^T\cdot \mathbf{p}^A}{|\mathbf{r}-\mathbf{R}^A|^3} - \frac{1}{2}\frac{\mathbf{r}^T\cdot \mathbf{D}^A\cdot\mathbf{r}}{|\mathbf{r}-\mathbf{R}^A|^5} \;,
 \label{eq:potential_from_analytical_formula}
\end{equation}
where the relations between $\mathbf p$ and $\mathbf D$ on one hand and the $Q_{\ell m} (\ell=1,2)$ on the other hand are detailed in Appendix~\ref{sec:Definitions and relations between the multipole moments}.

From the above expressions the electrostatic potential of the fragment can be calculated in terms of the decomposition of its atom-centered
potentials, i.e.
\begin{equation}
 V^\mathfrak F(\mathbf{r}) = \sum_{A \in \mathfrak F} V^A(\mathbf{r}-\mathbf r_A) \;.
\end{equation}
In a QM/MM approach using an electrostatic or even polarized embedding~\cite{bakowies-hybrid-1996,lin-QM/MM-2007,senn-QM/MM-2009},
the potential $V^\mathfrak F(\mathbf r)$ can then be added to the QM Hamiltonian as an additional external potential.

\subsection{Choice of the diffusive centers}
\label{sec: Centers of the multipoles}
As the reference quantities in our method are the fragment (and not the atomic) multipoles, there might be other representations of $\rho^\mathfrak F$ different from the above which would provide the same fragment multipoles, but a \emph{different} electrostatic potential. In other terms, as the fragment is in principle \emph{not} immediately separable in terms of its atomic contributions, other centers might be chosen instead of the atomic positions.
This fact has to be taken into account if one is interested, for instance, in using such a method to express accurate short-range representations of $V^\mathfrak F$, like for example electrostatic potential fitting methods. 

However, we nevertheless also want to assess the correctness of this approximation closer to the atoms.
Even though we do not perform potential fitting or use other similar techniques and thus only expect a qualitative agreement, we show in Tab.~\ref{tab:molecule_potential_difference} a quantitative comparison for three small molecules (\ce{H2O}, \ce{HF}, \ce{H2}), namely the values of 
\begin{equation}
 \Delta V = \frac{\int (V_\mathrm{exact}(\mathbf{r})-V_\mathrm{multipoles}(\mathbf{r}))^2 \, \mathrm{d}\mathbf{r}}{\int V_\mathrm{exact}(\mathbf{r})^2 \,\mathrm{d}\mathbf{r}} \;,
 \label{eq:relative_error_potential}
\end{equation}
where $V_\mathrm{exact}(\mathbf{r})$ is the exact electrostatic potential and $V_\mathrm{multipoles}(\mathbf{r})$ its approximation using the multipole expansion.
Since the multipoles are only expected to yield a reasonable representation of the electrostatic potential far away from the atoms, we report this difference for various threshold values,
meaning that only grid points where the charge density is below this given threshold (and thus far away from the molecule) contribute to the integrals. 

As can be seen, for \ce{H2O} and \ce{HF} the agreement between the exact and approximate potentials is already good rather close to the atoms (relative deviations of a few percent) and indeed becomes even better the smaller the aforementioned threshold is.
However, for \ce{H2} the results are considerably worse, which is a consequence of the fact that we place our atomic multipoles exclusively on the atoms and not, for instance, on covalent bonds.
Both \ce{H2O} and \ce{HF} have a strong dipole, resulting in considerable atomic partial charges, and the electrostatic potential can thus be well represented by placing the multipoles onto the atoms.

For \ce{H2}, on the other hand, the molecular dipole is zero due to symmetry, and the atomic partial charges are thus zero as well --- the charge is rather localized \textit{between} the atoms. However, since we force the multipoles to be located \textit{on} the atoms, the approximation of the electrostatic potential shows a considerable deviation from the exact one.

\begin{figure}
 \includegraphics[width=0.5\textwidth]{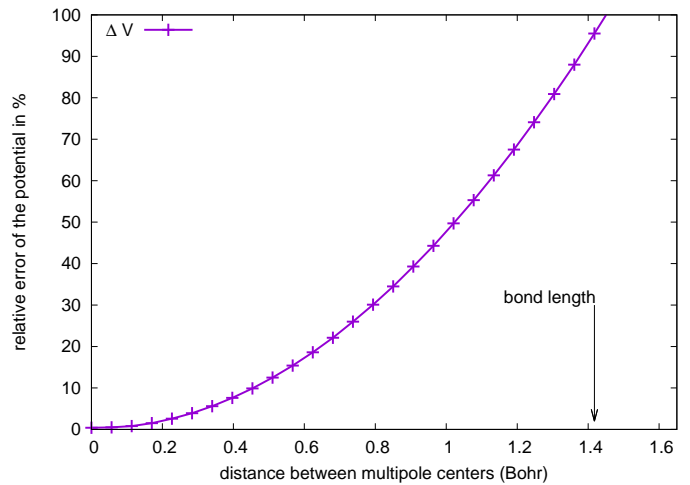}
 \caption{Relative error of the multipole potential for \ce{H2} (according to Eq.~\eqref{eq:relative_error_potential}) as a function of the distance between the multipole centers.
 The centers are always symmetric with respect to the bond center, i.e.\ a distance of 0 corresponds to a setup where both multipoles are located at the center.}
 \label{H2_pot}
\end{figure}

This problem can be solved by providing additional flexibility with respect to the choice of the multipole centers.
In Fig.~\ref{H2_pot} we show the error of Eq.~\eqref{eq:relative_error_potential} for \ce{H2} as a function of the distance between the two multipole centers, keeping them always symmetric with respect to the bond center.
It is important to stress that we only move the multipoles corresponding to electronic charge, whereas the monopoles corresponding to the positive counter charge of the nuclei remain at their original positions.
As can be seen the error decreases as the electronic centers come closer to each other, and becomes virtually zero as soon as they are overlapping each other in the middle of the bond.
The exact values for this last case are also noted in Tab.~\ref{tab:molecule_potential_difference}.

\section{Illustrative Applications}
\label{sec: applications}
In the following we show various applications of our approach for the effective electrostatic embedding of an active QM region into the potential generated by a set of fragment multipoles.
Using the purity indicator of Eq.~\eqref{eq:normalized_purity_indicator} we make sure that the fragments are properly chosen and the associated multipoles can thus be considered as reliable pseudo-observables.
With respect to the projector matrix $\mathbf{R}^\mathfrak{F}$, we always use the Mulliken method. 

In all examples, the density matrix is described by a \emph{minimal} set of \emph{localized in-situ optimized} basis functions,
as implemented in the DFT code \textsc{BigDFT}~\cite{genovese-daubechies-2008,mohr-daubechies-2014,mohr-accurate-2015}.
The use of a minimal set helps in the proper identification of the fragment, whereas the in-situ optimization nevertheless guarantees an accurate description of the electronic structure~\cite{mohr-fragments1-2017}.
However, as discussed in detail in Ref.~\citenum{mohr-fragments1-2017}, by choosing a suitable population analysis
--- i.e.\ yielding a low value for the so-called purity indicator --- 
our methodology can also be applied to calculations with other localized functions, e.g.\ fixed atomic orbitals or Gaussians.
Unless otherwise stated, the PBE functional was used.

\subsection{Ordered system replica --- the case of bulk pentacene}
\label{sec: pentacene}

\begin{figure}
 \includegraphics[width=1.0\columnwidth]{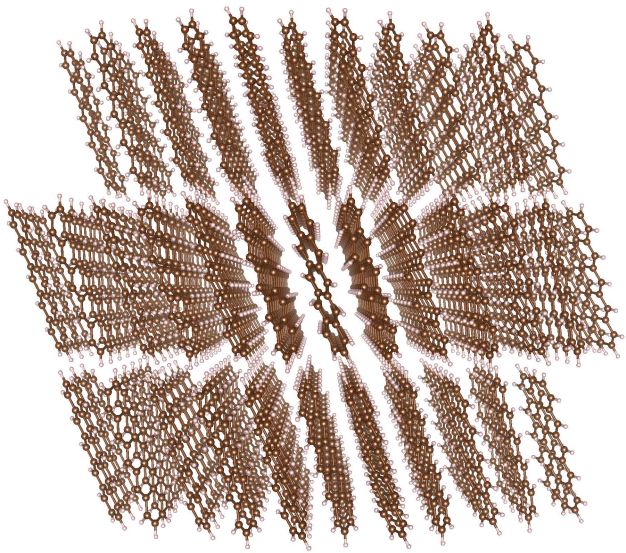}
 \caption{Visualization~\cite{momma-VESTA-2011} of the assembled pentacene molecules.}
 \label{fig:pentacene_large_rotated}
\end{figure}

As a first example we take (a portion of) bulk pentacene (PEN), as shown in Fig.~\ref{fig:pentacene_large_rotated}.
A single pentacene molecule has neither a net monopole nor dipole, 
so that the first non-zero contribution to the electrostatic potential should come from the rather short-range quadrupole term.
There is thus no ambiguity in the identification of the fragment,
as the presence of the neighboring molecules introduces only band effects and negligible hybridization among the PEN KS orbitals.
This system therefore represents an interesting playground for a basic QM/QM description; we use here the term QM/QM (instead of QM/MM) since all quantities are extracted from QM calculations.

\paragraph{Validity of the fragment approximation}
First we performed a full QM calculation of the entire assembly of pentacenes, consisting in total of 191 molecules (6876 atoms), which will serve as a reference.
The binding energy of the pentacene molecules is found to be \unit[282]{meV} per fragment unit,
confirming our assumption that there is little direct interaction between the pentacenes, and that one pentacene can be considered a well defined fragment of the entire system. We observe in the bulk system, thanks to the formation of bands, a relatively small  reduction (\unit[20]{meV}) in the HOMO-LUMO energy gap (see Tab.~\ref{tab:energies_different_functionals}), expressed as the peak-to-peak distance of the density of states (DoS).

To analyze such a subsystem identification in a more quantitative way, we looked at the purity indicator $\Pi$ as defined in Eq.~\eqref{eq:normalized_purity_indicator}.
Its value for each of the single pentacene molecules is of the order of $3 \cdot 10^{-3}$, which confirms that it is more than reasonable to consider a single pentacene molecule as a well-defined fragment of the system.
\begin{table*}\tablesize
 \begin{tabular}{l l l rrrr r rrrr}
  \toprule
    &&& \multicolumn{4}{c}{embedded} && \multicolumn{4}{c}{gas phase} \\
    \cmidrule{4-7} \cmidrule{9-12} \\
    &&& \multicolumn{1}{c}{LDA} &  \multicolumn{1}{c}{PBE} &  \multicolumn{1}{c}{PBE0} & \multicolumn{1}{c}{B3LYP} && \multicolumn{1}{c}{LDA} &  \multicolumn{1}{c}{PBE} &  \multicolumn{1}{c}{PBE0} & \multicolumn{1}{c}{B3LYP} \\
       \cmidrule{4-7} \cmidrule{9-12} \\
  \cmidrule{1-12}
    \multirow{3}{*}{1 PEN} 
  & $E_{HOMO}$  && -4.75 & -4.59 & -5.24 & -5.02 && -4.63 & -4.47 & -5.12 & -4.90 \\ 
  & $E_{LUMO}$  && -3.52 & -3.33 & -2.72 & -2.76 && -3.39 & -3.20 & -2.59 & -2.63 \\
  & $E_{gap}$   &&  1.23 &  1.26 &  2.52 &  2.26 &&  1.24 &  1.27 &  2.53 &  2.27 \\
   \cmidrule{1-12}
   \multirow{3}{*}{3 PEN} 
   & $\langle E_{HOMO} \rangle$  && $-4.38(18)$ & $-4.21(18)$ & $-4.83(20)$ & $-4.62(19)$ && $-4.51(22)$ & $-4.33(22)$ & $-4.97(24)$ & $-4.75(23)$ \\
   & $\langle E_{LUMO} \rangle$  && $-3.21(22)$ & $-3.02(22)$ & $-2.40(25)$ & $-2.44(24)$ && $-3.31(24)$ & $-3.11(24)$ & $-2.49(27)$ & $-2.53(26)$ \\
   & $\langle E_{gap} \rangle$   && $ 1.17(38)$ & $ 1.19(38)$ & $ 2.43(42)$ & $ 2.18(40)$ && $ 1.20(41)$ & $ 1.22(41)$ & $ 2.47(46)$ & $ 2.22(44)$ \\
   \cmidrule{1-12}
   \multirow{3}{*}{full QM} 
   & $\langle E_{HOMO} \rangle$  && --- & --- & --- & --- && $-4.09$ & $-3.88$ & --- & --- \\
   & $\langle E_{LUMO} \rangle$  && --- & --- & --- & --- && $-2.88$ & $-2.64$ & --- & --- \\
   & $\langle E_{gap} \rangle$   && --- & --- & --- & --- && $ 1.21$ & $ 1.25$ & --- & --- \\
 \bottomrule
 \end{tabular}
 
 \caption{HOMO level, LUMO level and HOMO-LUMO gap in eV for one and three pentacenes, both embedded and isolated, for various functionals. For three pentacenes, we consider, instead of the HOMO and LUMO levels, the average value of $\{\mathrm{E_{HOMO-2}, E_{HOMO-1}, E_{HOMO}}\}$ and $\{\mathrm{E_{LUMO}, E_{LUMO+1}, E_{LUMO+2}}\}$, respectively, in order to account for band effects; in parentheses we also give the standard deviation. Additionally we also show the values for the full QM bulk calculation, where the HOMO and LUMO energies correspond to the top of the two smeared out HOMO and LUMO peaks in Fig.~\ref{fig:DoS_pentacene_QMMM}, respectively.
 }
\label{tab:energies_different_functionals}
\end{table*}

\begin{figure}
 \includegraphics[width=1.0\columnwidth]{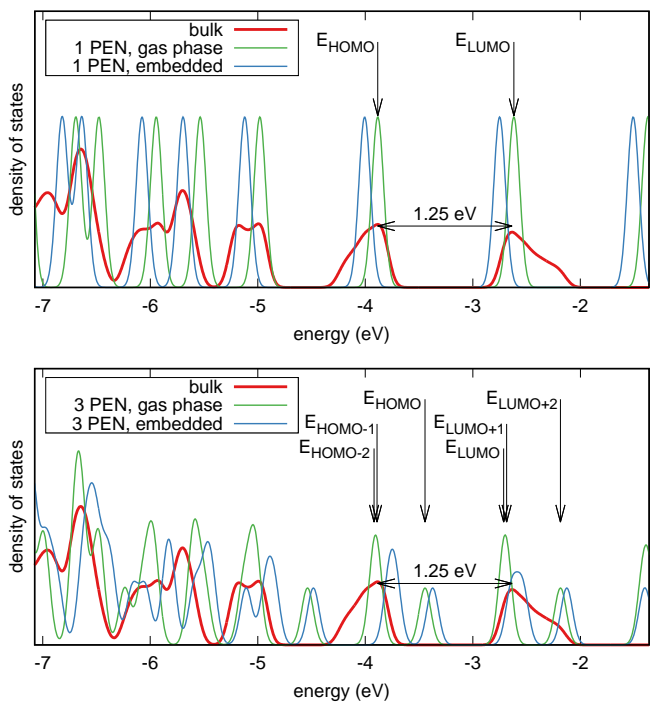}
 \caption{Comparison of the density of states for 1 (upper panel) and 3 (lower panel) pentacenes (calculated with the cubic version of \textsc{BigDFT}), together with the density of states of the entire system (calculated with the linear version of \textsc{BigDFT}). For easier comparison, all curves have been normalized by dividing by the number of atoms and shifted by $\Delta E$, where $\Delta E$ is defined such that the HOMO level of 1 PEN gas phase coincides with the HOMO peak of the full QM bulk calculation.
 The overall electronic structure for all three setups is very similar.
 Additionally we also show the HOMO-LUMO gap of the full QM calculation, which is \unit[1.25]{eV}.
 For all curves a Gaussian smearing with $\sigma=\unit[0.05]{eV}$ was applied.
}
  \label{fig:DoS_pentacene_QMMM}
 \end{figure}

\paragraph{Embedding of central pentacene molecules}
To validate a QM/QM approach in this system, we then chose a few central molecules and embedded them in the multipole potential 
generated by the other PEN molecules, by averaging the molecular multipoles
found in the reference full QM calculation.
In order to generate the proper bulk environment, surface PEN have not been considered for the averaging.
In this way we were able to reproduce the electronic structure of the full QM results, as shown in Fig.~\ref{fig:DoS_pentacene_QMMM}, for both one and three pentacenes in the QM region.
This similarity between the full QM calculation for the bulk on the one hand and the gas phase and embedded ones for the individual PENs on the other hand confirms 
our initial assumption that such an electrostatic embedding should work for this system, as the fragments can be easily defined and there is only little interaction between them.
Clearly the HOMO and LUMO levels split up when going from one to three pentacenes; this effect would continue for even more pentacenes, eventually forming the bands that we see for the bulk calculation. To take into account band effects we considered for the calculation of the gaps the ``average HOMO/LUMO energies''; for one pentacene this corresponds to the ordinary HOMO and LUMO states, whereas for 3 pentacenes it corresponds to the mean of the values $\{\mathrm{HOMO-2,HOMO-1,HOMO}\}$ and $\{\mathrm{LUMO,LUMO+1,LUMO+2}\}$, respectively.
Like for the complete bulk calculations, we observe that also these values are slightly lower in embedded phase than the corresponding quantities in gas phase.

\paragraph{Embedding with different functionals}
Now that we have validated the electrostatic embedding for this system --- by showing that the electronic structure of the large system can be represented well by taking only one molecule and using the 
multipoles to generate the external potential --- we can use this technique to calculate quantities which are more easily -- if not exclusively -- accessible within such a reduced complexity setup.
As an example, we can use more expensive functionals for the QM active region, and in this way perform calculations using hybrid functionals, which would be computationally extremely expensive --- or even out of reach ---
for the full QM system containing several thousand atoms.
We show in Tab.~\ref{tab:energies_different_functionals} the values of the HOMO-LUMO gap for one and three pentacenes,
taking as illustrations the popular LDA~\cite{goedecker-separable-1994}, PBE~\cite{perdew-generalized-1996}, PBE0~\cite{adamo-toward-1999,ernzerhof-assessment-1999} and B3LYP~\cite{stephens-ab-initio-1994} functionals,
as implemented in the \textsc{Libxc} library~\cite{marques-2012-libxc}.

As can be seen, there is only a small difference --- still corresponding to a lowering of the gap by \unit[20]{meV} --- between the gas phase and the embedded calculations, again confirming that the PEN molecules represent ideal fragments with only little interaction among each other. However, and more importantly, the hybrid functionals yield considerably larger values for the gap than LDA and PBE and thus correct the well-known gap underestimation of these (semi-)local functionals~\cite{perdew-density-1985}.
Such an embedded approach can thus pave a way
towards the routine usage of expensive QM descriptions
also for large systems, where a straightforward calculation would be out of reach.
Moreover such an embedding can also be used to couple DFT to even more accurate wave function methods~\cite{bennie-2016-a-projector}.

\subsection{Disordered and heterogeneous systems --- the case of DNA in water}
\label{sec: DNA}
\begin{figure*}
 \includegraphics[width=0.9\textwidth]{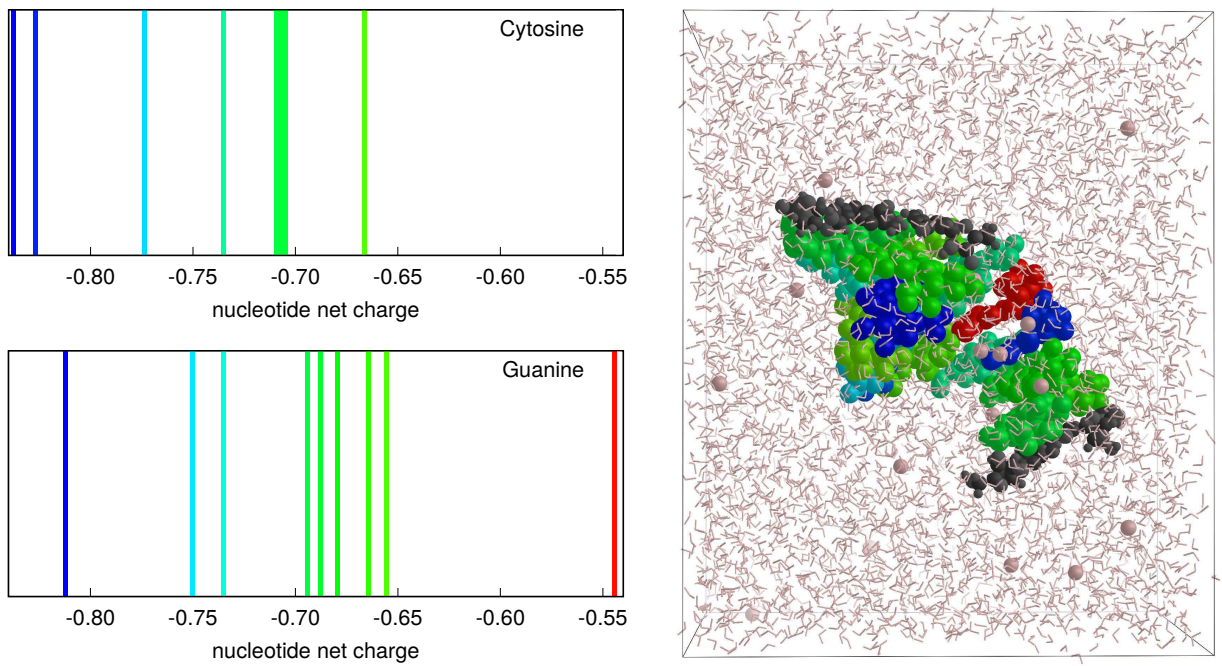}
 \caption{Visualization of the solvated DNA. The color code represents the charges of the DNA residues as found from the full QM calculation with the method described in Ref.~\citenum{mohr-fragments1-2017}. The terminal nucleotides do not represent perfect Cyt and Gua nucleotides any more and are thus neither listed in the histogram nor colored in the figure. The solvent water molecules and Na ions are represented in faded pink.}
 \label{fig:dnainsolvent}
\end{figure*}

We have shown in Sec.~\ref{sec: pentacene}
the validation and benefits of the electrostatic embedding for a homogeneous system where the fragments can easily be identified by chemical intuition.
Now we want to apply this formalism to a more complicated system, namely the case of solvated DNA. 
In this setup the interesting part --- the DNA --- is rather heterogeneous, but the solvent part might be a good candidate for our fragment multipole scheme.
As a specific example we take a snapshot from an MD simulation --- run with Amber 11~\cite{case-the-2005,Amber11} and the ff99SB force field~\cite{hornak-comparison-2006} --- for a 11 base pair DNA fragment (made only of Guanine and Cytosine nucleotides) being embedded into a sodium-water solution, giving in total 15,613 atoms.
The system is depicted in Fig.~\ref{fig:dnainsolvent}.


\subsubsection{Necessity of explicit solvation}
It is known that important quantities of biological systems -- such as the HOMO-LUMO gap --- are very sensitive to a proper setup of the electrostatic environment~\cite{Lever-2013}. 
However, as is shown in Ref.~\citenum{mohr-fragments1-2017}, the DNA environment constituents --- 
i.e.\ the \ce{Na} atoms and the water molecules --- exhibit quite low values for the purity indicator of Eq.~\eqref{eq:normalized_purity_indicator} and can 
thus be considered as well-defined fragments.
Consequently we might nevertheless ask how much the electrostatic environment they generate actually influences
the properties of the DNA and to what extent it is important to include many explicit solvent atoms in the calculation, i.e.\ how big the active QM region should be~\cite{kulik-how-2016}.
To shed light on this, we present in Tab.~\ref{tab:non-purity_shells} the purity indicator $\Pi$ from Eq.~\eqref{eq:normalized_purity_indicator} for the DNA plus various solvent shells from \unit[0]{\r{A}} to \unit[8]{\r{A}}.
All values are very small, indicating that all setups represent sensible fragment choices.
Unsurprisingly, the values become even smaller the larger the shell is; however, 
there is a clear drop when going from \unit[2]{\r{A}} to \unit[4]{\r{A}},
indicating that the interaction of the solvated DNA with its environment decreases considerably at that size.

\begin{table}\tablesize
  \begin{tabular}{l l rrrrr}
  \toprule
   solvent shell  && \multicolumn{1}{c}{\unit[0]{\r{A}}} & \multicolumn{1}{c}{\unit[2]{\r{A}}} & \multicolumn{1}{c}{\unit[4]{\r{A}}} & \multicolumn{1}{c}{\unit[6]{\r{A}}} & \multicolumn{1}{c}{\unit[8]{\r{A}}} \\
   \cmidrule{3-7}
    number of shell atoms &&    0 &   45 & 1164 & 2210 & 3390 \\
    purity indicator $\times 10^3$ && $5.0 $ & $4.9$ & $2.8$ & $2.2$ & $1.7 $ \\
    quality && \multicolumn{1}{c}{\good} & \multicolumn{1}{c}{\good} & \multicolumn{1}{c}{\good} & \multicolumn{1}{c}{\good} & \multicolumn{1}{c}{\good} \\ 
  \bottomrule
  \end{tabular}
  \caption{
  Purity indicator according to Eq.~\eqref{eq:normalized_purity_indicator}, within the Mulliken population analysis scheme,
    for the DNA plus various shells of solvent, from \unit[0]{\r{A}} (i.e.\ the pure DNA) up to \unit[8]{\r{A}}. 
 The values have been multiplied by a factor of one thousand.
  }
  \label{tab:non-purity_shells}
\end{table}

\paragraph{References from complete QM calculations}
\label{ExplicitGP}
In order to further investigate the influence of the explicit solvent on the internal QM region,
we analyzed the partial charges and dipole moments of the DNA nucleotides as a function of the quantity of surrounding solvent, doing calculations for solvation shells of \unit[0]{\r{A}}, \unit[2]{\r{A}}, \unit[4]{\r{A}} and \unit[6]{\r{A}}.
As a reference we took a full QM calculation of the entire system, performed with the linear scaling approach of \textsc{BigDFT}~\cite{mohr-daubechies-2014,mohr-accurate-2015}; from this calculation we also extracted the set of atomic multipoles which will be used in the subsequent steps.
The values of the charges for each of the nucleotides of the DNA in solution are depicted in Fig.~\ref{fig:dnainsolvent}.
This charge analysis also allows us to determine how much of the \ce{Na} charge has gone to the DNA. 
The 20 \ce{Na} atoms have lost 19.2 electrons (corresponding to an average ionization of 0.96), 
out of which 3.6 have gone to the water and the other 15.6 to the DNA.

\begin{figure*}
  \subfloat[][neutral]{
  \includegraphics[width=0.286\textwidth]{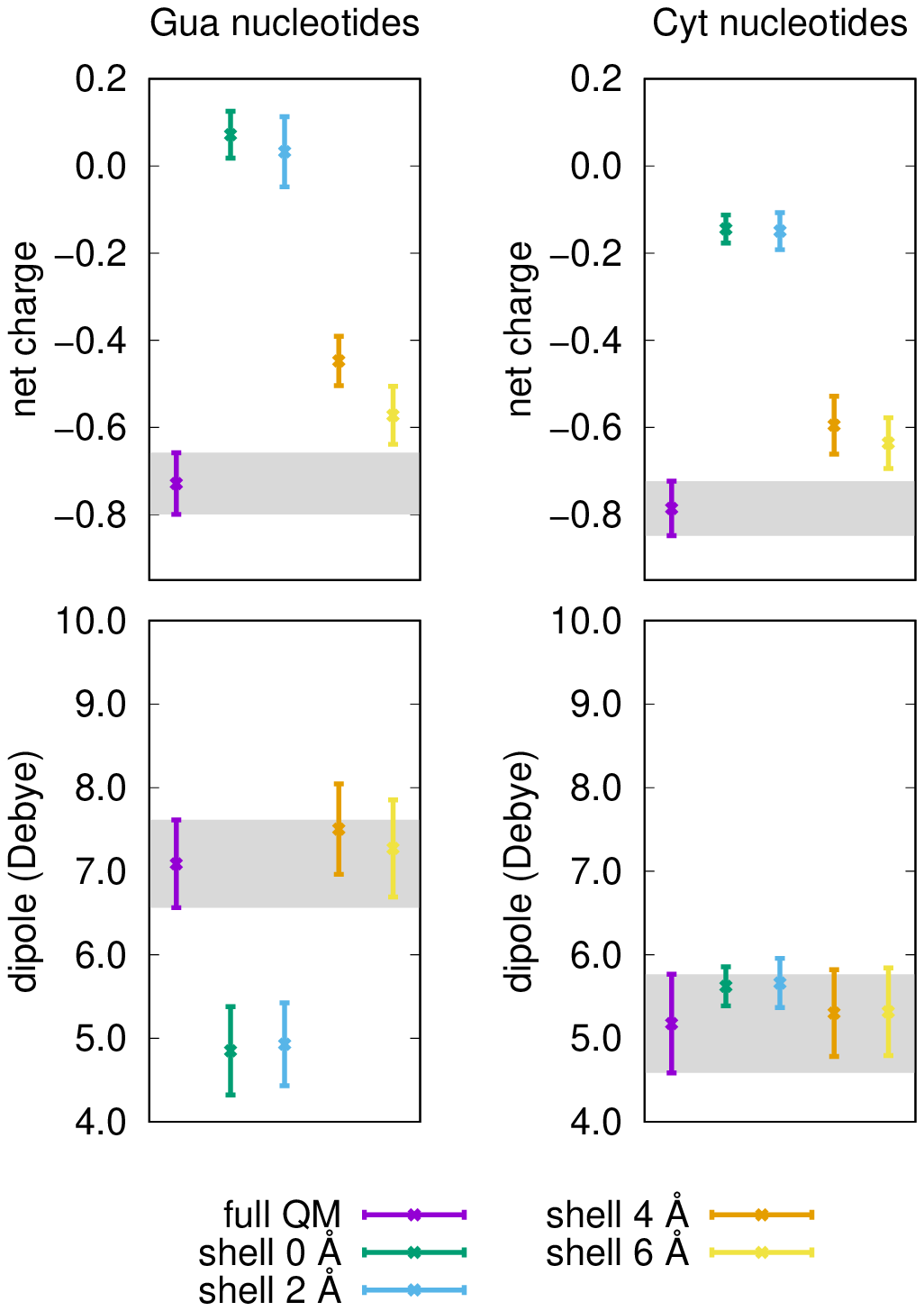}
  \label{fig:nucleotides_neutral_average}
 }
\qquad
 \subfloat[][charged]{
  \includegraphics[width=0.286\textwidth]{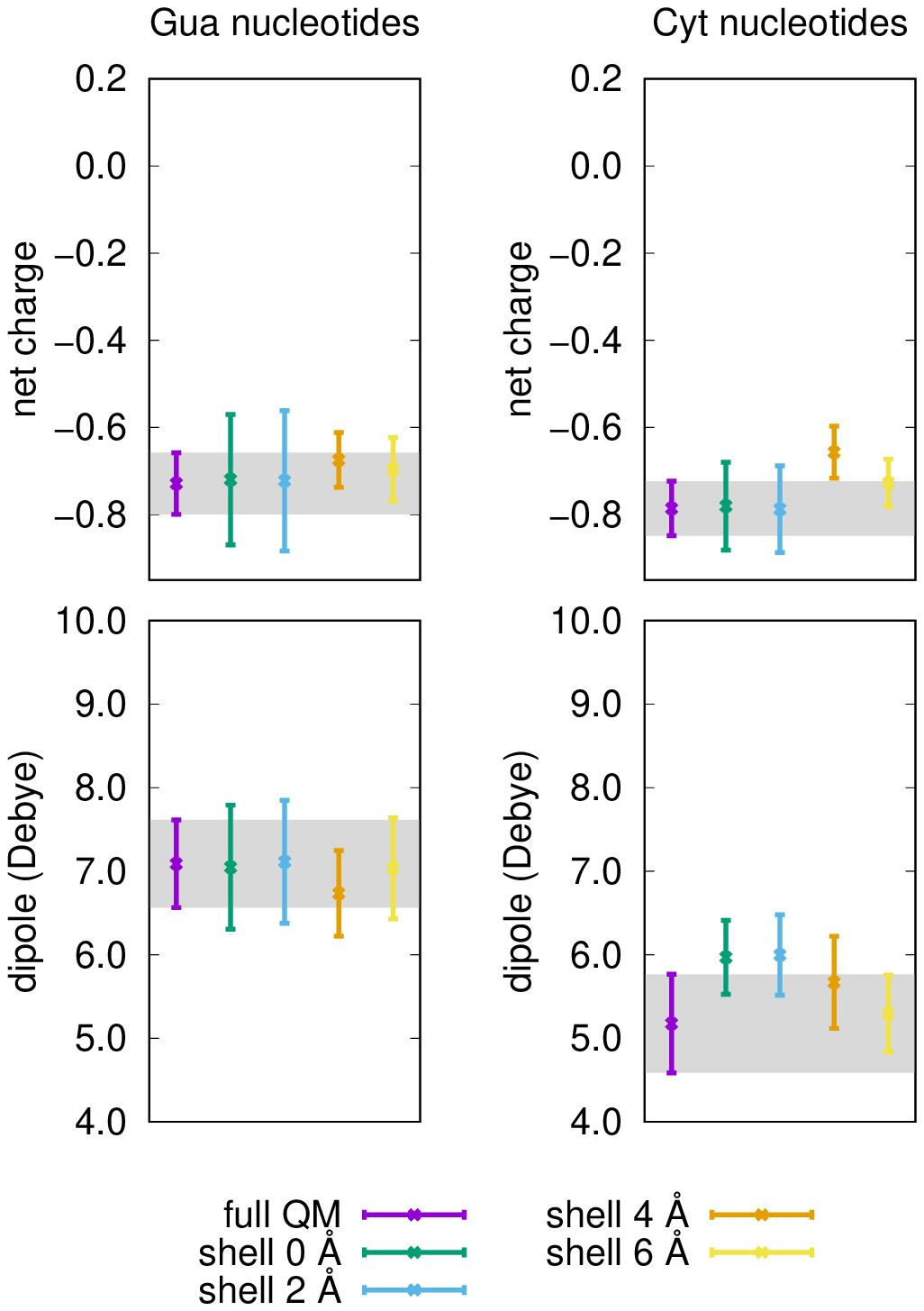}
  \label{fig:nucleotides_charged_average}
 }
\qquad
 \subfloat[][QM/QM]{
  \includegraphics[width=0.286\textwidth]{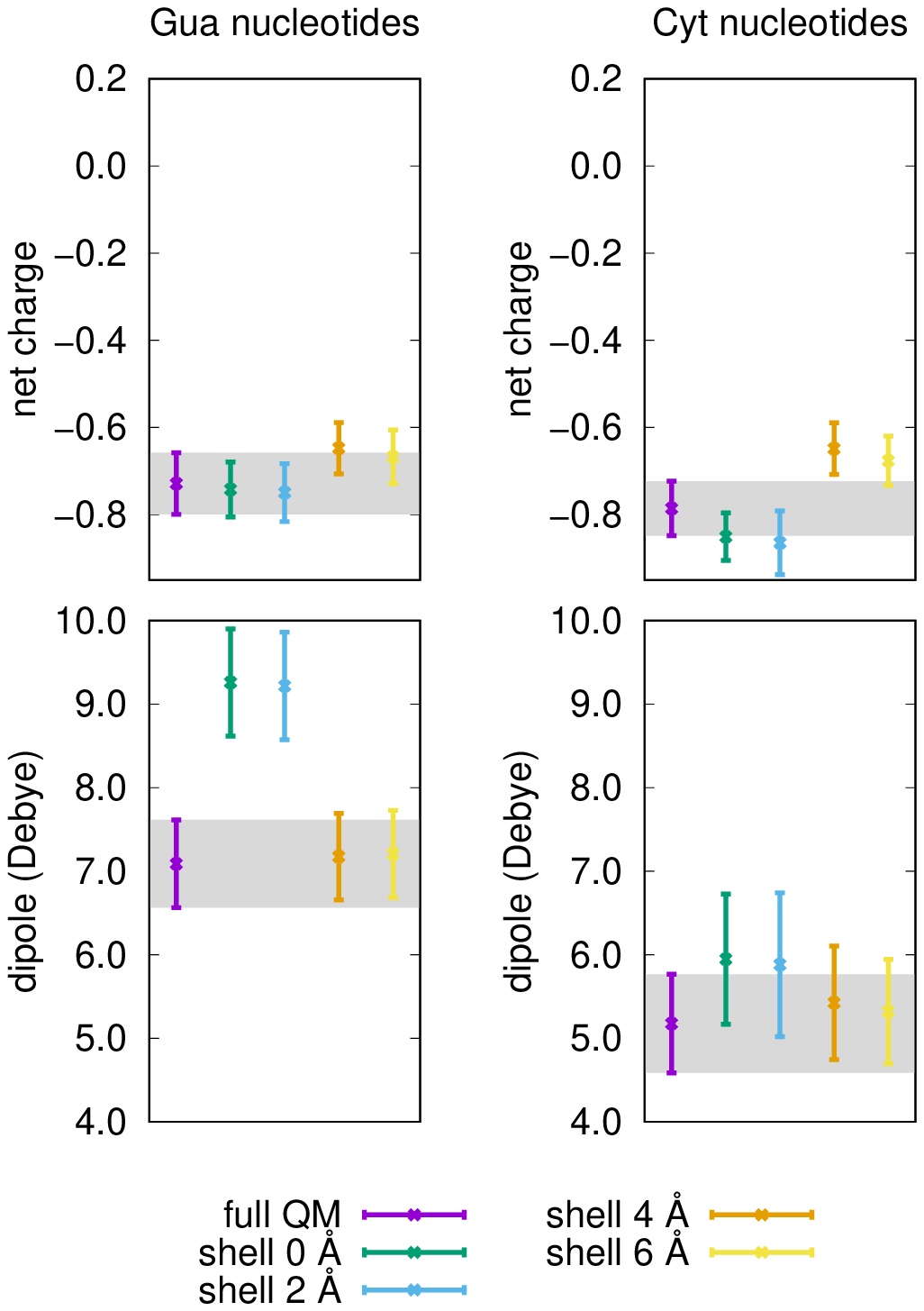}
  \label{fig:nucleotides_QMQM_average}
 }
 \caption{Mean value and standard deviation of the nucleotide charges and dipole moments for the neutral (\ref{fig:nucleotides_neutral_average}), charged (\ref{fig:nucleotides_charged_average}) and QM/QM (\ref{fig:nucleotides_QMQM_average}) setup. The standard deviation must not be interpreted as an error bar, but rather as the natural variation of the charges and dipoles among the nucleotides. Thus not only the mean value, but also the standard deviation should coincide with the reference full QM calculation.}
 \label{fig:nucleotides_average}
\end{figure*}

\paragraph{Test of non-embedded setups}
To test the actual influence that the environmental molecules have on the 
DNA region we first removed the solvent molecules outside of the explicit solvent shell and conducted calculations with two different setups.
i) In the ``neutral'' setup we 
performed a neutral calculation of the system as-is, knowing that this can be problematic --- in particular for small 
solvent shells --- as the DNA attracts charge from its environment.
ii) In the ``charged'' setup we explicitly charged the system with 
the negative countercharge of those atoms which are neglected with respect to the full QM calculation, ranging from 15.6 electrons for the naked DNA up to 8.9 electrons for a shell of \unit[6]{\r{A}}.
In Fig.~\ref{fig:nucleotides_average} we show the mean value and standard deviation for the nucleotide charges and dipole moments, with Fig.~\ref{fig:nucleotides_neutral_average} showing the neutral setup
and Fig.~\ref{fig:nucleotides_charged_average} the charged setup.
The results for the second setup are considerably more accurate, in particular for small solvent shells.
Nevertheless we also need in this case a shell of at least \unit[4]{\r{A}} to get reasonably close to the reference calculation where the entire solvent is explicitly taken into account.
This is again in remarkable correlation with the results of Tab.~\ref{tab:non-purity_shells}.

\begin{figure}
    \includegraphics[width=1.0\columnwidth]{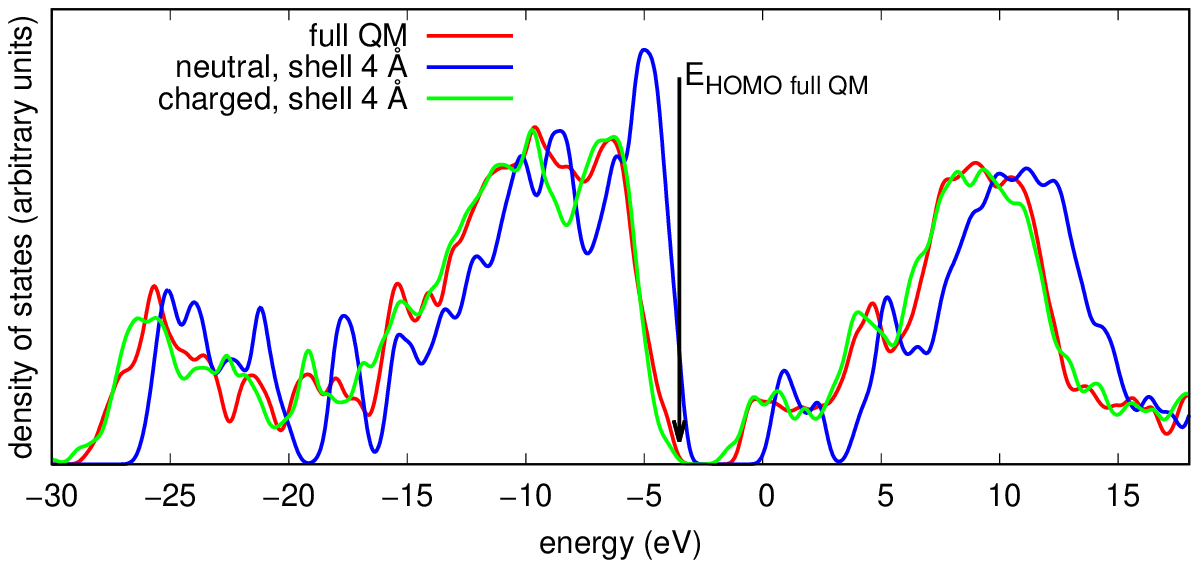}
    \caption{Partial density of states for the DNA subsystem, for the different setups described in the text. The curves have been shifted such that the HOMO energies coincide,
    and a Gaussian smearing with $\sigma=\unit[0.27]{eV}$ was applied.}
    \label{fig:DNA-water_pdos_multipoles}
   \end{figure}

In Fig.~\ref{fig:DNA-water_pdos_multipoles} we show the density of states of these two simplified QM setups, 
together with the partial density of states (PDoS) $f^\mathfrak{F}(\epsilon)$ for the DNA subsystem coming from the full QM calculation, which is given by
\begin{equation}
 f^\mathfrak{F}(\epsilon) = \sum_j \Tr \left(F \hat W^j \hat W^\mathfrak{F} \right) \delta(\epsilon-\epsilon_j) \;,
\end{equation}
where $\epsilon_j$ is the $j$th Kohn-Sham eigenvalue of the system, 
$\hat W^j$ is the projector selecting the $j$th KS orbital,
\begin{equation}
 \hat W^j= \ket{\psi_j}\bra{\psi_j} \;,
\end{equation}
and $\hat W^\mathfrak{F}$ is the Mulliken definition of the fragment.
Again we can see a very good agreement between the reference result and the charged QM setup with the \unit[4]{\r{A}}
explicit solvent shell.
   
 However a big difference becomes visible if we compare the absolute energy eigenvalues.
In Tab.~\ref{tab:DNA_eigenvalues} we show the HOMO energies and the value of the HOMO-LUMO gap for the different setups.
The gap for the neutral setup is basically zero --- a fact which is not visible from Fig.~\ref{fig:DNA-water_pdos_multipoles} due to the applied smearing.
This wrong description is due to the missing electrons which should be added to populate 
the states filled by the electrons that the nucleotides attract from the environment, and indeed the charged 
setup exhibits a more reasonable value for the gap.
On the other hand, the HOMO energy of the charged setup is largely positive, showing that
this state is not bound, and therefore it has no actual physical meaning for gas phase boundary conditions. 
Additionally the total energy of the charged setup is slightly higher (\unit[1.4]{eV}) than that of the neutral setup, also demonstrating that charging the system without a proper confining environment is not ideal.

\subsubsection{QM/QM embedding}
\label{sec: DNA in water embedded}
The comparisons with results obtained from a full QM calculation indicate that 
we have to somehow consider the environment in order to identify the ideal setup for a reduced complexity QM/QM calculation.
Therefore, we now want to add the embedding potential, calculated from the set of QM multipoles, to see whether this gives us a description which is of equal quality 
as the full QM calculation.

In Fig.~\ref{fig:nucleotides_QMQM_average} we show again the mean value and standard deviation of the nucleotide charges and dipoles, but this time using this embedded QM/QM approach;
again we charged the DNA with the negative countercharge of the atoms outside of the QM region.
The results are similar to the ``charged'' setup, and once more confirm the need for an explicit inclusion of an explicit solvent shell of at least \unit[4]{\r{A}} for this system.
For smaller radii of the explicit solvent shell, the well-known~\cite{senn-QM/MM-2009,neugebauer-subsystem-2009} phenomenon of ``overpolarization'' occurs:
Some nucleotides acquire an extra dipole as the KS orbitals are partially attracted by the environmental molecules.
In the literature there exist various approaches to avoid this problem, a popular one being the use of a delocalized charge distribution for the environment atoms~\cite{eichinger-1999-a-hybrid,das-2002-optimization,biswas-2005-a-regularized} --- a property which is indeed fulfilled using our ansatz.
In our case the overpolarization phenomenon might simply be ascribed to the presence of too few explicit solvent molecules. Indeed, as the number of solvent molecules becomes slightly larger (above \unit[4]{\r A}), the problem completely disappears.

Also with respect to Tab.~\ref{tab:DNA_eigenvalues}, the situation using this QM/QM approach is considerably improved.
The HOMO value is now negative, showing that the associated KS orbital is confined by the environment, and
the gap is large and slightly overestimated, which is sound considering that some molecules have been excluded from the QM calculation, thereby reducing the smearing due to disorder.
\begin{table}\tablesize
 \begin{tabular}{l l rrrr}
 \toprule
       && \multicolumn{1}{c}{full QM} & \multicolumn{1}{c}{neutral} & \multicolumn{1}{c}{charged} & \multicolumn{1}{c}{QM/QM} \\
  \cmidrule{3-6}
  HOMO && -3.52 & -6.41 (\good) & 7.98 (\bad) & -4.81 (\good) \\
  gap  &&  2.76 &  0.006 (\bad)  & 1.89 (\good)&  3.26 (\good) \\
 \bottomrule
 \end{tabular}
 \caption{HOMO energies and HOMO-LUMO gaps for the solvated DNA. For the neutral, charged and QM/QM setup we used a solvent shell of \unit[4]{\r{A}}. All values are given in eV.}
 \label{tab:DNA_eigenvalues}
\end{table}

Overall, the QM/QM setup is thus the only one which shows in all aspects a behavior which is similar to that of the full QM setup and thus validates this approach.
This demonstrates that we can efficiently represent the electrostatic environment via the atomic multipoles and thus achieve a considerable reduction in the complexity of a DFT calculation on a system of several thousand atoms.

\section{Conclusions and outlook}
\label{sec: conclusions}

In this paper we presented a simple method for reducing the complexity of large QM systems.
To do so, we split the system into an active QM part and an environment part, and --- upon partitioning the latter into well defined fragments ---  
represent the environment by a set of fragment multipoles, resulting in an effective electrostatic embedding of the QM part.
The approach works best for fragment multipoles that can be considered as ``pseudo-observables'', which means that they are quantities with an interpretable physico-chemical meaning.
This property is closely related to a proper fragment definition, which can be easily verified in a quantitative way using our so-called purity indicator.
In order to get a reliable set of fragment multipoles it is advantageous to use a set of minimal and in-situ optimized basis functions;
in such a favorable situation the embedding exhibits a very high quality, and there is no need to perform any potential fitting or similar techniques, nor to deal with approaches describing explicitly the interface between the QM and the embedding region.

Since our approach is based on an \emph{a posteriori} identification of the fragments and the associated multipoles out of full QM calculations, it can easily be validated by comparing it with unbiased results stemming from such full QM setups.
Thanks to the fact that we have implemented both functionalities --- i.e.\ a linear scaling full QM calculation and the embedding approach --- in the same code, namely \textsc{BigDFT}, this verification of the complexity reduction can be done in a straightforward way.
Once this validation is done, the embedding can then be used for subsequent calculations within a reduced complexity scheme, 
giving access to quantities which would not be reachable otherwise, 
as for example demonstrated by the use of hybrid functionals which would considerably increase the computational cost
for a straightforward full QM calculation.

In this way this approach of fragment identification and representation allows one to considerably lower the number of degrees of freedom and thus to reduce the complexity of QM treatments of large systems.
When properly employed this simplification does not notably affect the accuracy of the description of a quantum mechanical system and keeps the essential physico-chemical properties unaffected.
In this sense our ansatz allows to couple various levels of theory,
and the efficient coarse graining of the QM description that we obtain allows in particular ab-initio calculations to be coupled with classical approaches. 
This scheme paves the way towards powerful QM/QM or QM/MM calculations for very large systems 
and is thus an important link within a multiscale approach aiming at the bridging of so-called time and lengthscale gaps~\cite{ratcliff-2016-challenges}.

\section{Acknowledgments}
We would like to thank Thierry Deutsch for valuable discussions and F\'atima Lucas for providing various test systems and helpful discussions. 
This research used resources of the Argonne Leadership Computing Facility, which is a DOE Office of Science User Facility supported under Contract DE-AC02-06CH11357.
SM acknowledges support from the European Centre of Excellence MaX (project ID 676598).
LG acknowledges support from the EU ExtMOS project (project ID 646176) and the European Centre of Excellence EoCoE (project ID 676629).

\appendix

\section{Approximating the electrostatic density from a set of multipoles}
\label{sec:Approximating the electrostatic density from a set of multipole}
Suppose we have extracted the multipole coefficients $Q_{\ell m}$, defined in Eq.~\eqref{basicqlm},
of a charge density $\rho(\mathbf r)$.
Clearly, the $Q_{\ell m}$ alone do not suffice to completely determine the original function $\rho$.
It can be easily seen that \textit{any} function of the form
\begin{equation}
 f(\mathbf r) = \sqrt{4\pi}\sum_{\ell,m} \sqrt{2\ell+1} Q_{\ell m} \phi_\ell(r) \frac{S_{\ell m}(\mathbf r)}{r^{2\ell}}
\end{equation}
yields the same multipoles as the original function $\rho$ as long as the radial functions $\phi_\ell(r)$ fulfill,
$\forall \ell$, the normalization condition
\begin{equation}\label{norm}
 \int_0^\infty \mathrm d r \phi_\ell(r) = \frac{1}{4 \pi} \;.
\end{equation}
When approximating a charge density we therefore have to \emph{choose} the family of functions $\phi_\ell(r)$
that express their radial behaviour.
It is convenient to employ functions of the form
\begin{equation}
 \phi_\ell(r) = \frac{a_\ell}{\sqrt{4 \pi (2 \ell +1)}} r^{2 \ell} \phi(r) \;,
 \label{eq:general_radial_function}
\end{equation}
where $\phi(r)$ is some regular radial function and $a_\ell$ a normalization coefficient ensuring the validity of Eq.~\eqref{norm}.
In this way we get for the reconstructed function
\begin{equation}
 f(\mathbf r) = \phi(r)\sum_{\ell,m} Q_{\ell m} a_\ell S_{\ell m}(\mathbf r) \;.
 \label{eq:function_from_multipoles}
\end{equation}
In our approach we use Gaussians for $\phi(r)$:
\begin{equation}
 \phi(r) = e^{-\frac{r^2}{2\sigma^2}} \; , 
\end{equation}
and the normalization coefficient $a_\ell$ is then given by
\begin{equation}
 1/a_\ell = \frac{\sigma^{2\ell+3} 2^{\ell+2} \sqrt{2}\pi}{\sqrt{4 \pi (2 \ell +1)} } \Gamma\left(\frac{3}{2} + \ell\right) \; .
\end{equation}

\section{Definitions and relations between the multipole moments}
\label{sec:Definitions and relations between the multipole moments}
There are two possible ways to define the monopole, dipole and quadrupole moments: Either using the basic formula of Eq.~\eqref{basicqlm}, or directly as 
\begin{equation}
 \begin{aligned}
  & q = \int \rho(\mathbf{r}) \, \mathrm{d}\mathbf{r} \;, \\
  & p_i = \int r_i \rho(\mathbf{r}) \, \mathrm{d}\mathbf{r} \;, \\
  & D_{ij}=\int (3r_ir_j-\mathbf{r}^2\delta_{ij})\rho(\mathbf{r}) \,\mathrm{d}\mathbf{r} \;.
 \end{aligned}
 \label{eq:direct_definition_multipoles}
\end{equation}
By comparing them, it follows that the monopole and dipole terms are identical up to a reordering:
\begin{equation}
 \begin{aligned}
  & q = Q_{00} \;, \\
  & \mathbf{p} = \begin{pmatrix} Q_{11}\\Q_{1-1}\\Q_{10} \end{pmatrix} \;.
 \end{aligned}
\end{equation}
For the quadrupoles we get for the off-diagonal elements
\begin{equation}
 \begin{aligned}
  & D_{12} = D_{21} = \sqrt{3}Q_{2-2} \;, \\
  & D_{13} = D_{31} = \sqrt{3}Q_{21} \;, \\
  & D_{23} = D_{32} = \sqrt{3}Q_{2-1} \;.
 \end{aligned}
\end{equation}
and for the diagonal elements (using the fact that $\mathbf{D}$ is traceless)
\begin{equation}
 \begin{aligned}
  & D_{33} = 2Q_{20} \;, \\
  & D_{11}-D_{22} = \sqrt{\frac{4}{3}}Q_{22} \;, \\
  & D_{11} + D_{22} + D_{33} = 0 \;.
 \end{aligned}
\end{equation}
This gives rise to the linear system
\begin{equation}
 \begin{pmatrix} 0&0&1\\1&-1&0\\1&1&1 \end{pmatrix} \begin{pmatrix} D_{11}\\D_{22}\\D_{33} \end{pmatrix} = \sqrt{\frac{4}{3}}\begin{pmatrix} \sqrt{3}Q_{20}\\Q_{22}\\0 \end{pmatrix} \;,
\end{equation}
whose solution leads to final result
\begin{multline}
 \mathbf{D}=\frac{1}{\sqrt{3}} \times \\ \begin{pmatrix} -\sqrt{3}Q_{20}+Q_{22} & 3Q_{2-2} & 3Q_{21} \\ 3Q_{2-2} & -\sqrt{3}Q_{20}-Q_{22} & 3Q_{2-1} \\ 3Q_{21} & 3Q_{2-1} & 2\sqrt{3}Q_{20} \end{pmatrix} \;.
\end{multline}

\bibliography{citationlist}

\end{document}